\documentclass{iopart}
\pdfoutput=1

\usepackage{iopams} 
\expandafter\let\csname equation*\endcsname\relax
\expandafter\let\csname endequation*\endcsname\relax 
\usepackage{amsmath}
\usepackage{amssymb}
\usepackage{graphicx}
\usepackage{subfig}
\usepackage{subfloat}
\usepackage{color}
\usepackage{multirow}

\begin{document}

\title[]{Dependence structure of market states}

\author{Desislava Chetalova, Marcel Wollschl{\"a}ger and Rudi Sch\"afer}

\address{Fakult\"at f\"ur Physik, Universit\"at Duisburg--Essen, D--47048 Duisburg, Germany}
\ead{desislava.chetalova@uni-due.de, marcel.wollschlaeger@uni-due.de, rudi.schaefer@uni-due.de}

\begin{abstract}
We study the dependence structure of market states by estimating empirical pairwise copulas of daily stock returns. 
We consider both original returns, which exhibit time-varying trends and volatilities, as well as locally normalized ones, where the non-stationarity has been removed.
The empirical pairwise copula for each state is compared with a bivariate K-copula.
This copula arises from a recently introduced random matrix model, in which non-stationary correlations between returns are modeled by an ensemble of random matrices. 
The comparison reveals overall good agreement between empirical and analytical copulas, especially for locally normalized returns. Still, there are some deviations in the tails.
Furthermore, we find an asymmetry in the dependence structure of market states. The empirical pairwise copulas exhibit a stronger lower tail dependence, particularly in times of crisis.
\end{abstract}

\maketitle

\section{Introduction}    \label{section1}

The concept of copulas was introduced by Sklar in 1959 \cite{Sklar1959, Sklar1973} to study the linkage between multivariate distribution functions and their univariate marginals. 
Since then, copulas have gained growing importance as a tool for modeling statistical dependence of random variables in many fields. In finance, the usage of copulas is relatively 
new but it has already found application in risk management, see e.g., \cite{Embrechts2002, Embrechts2003, McNeil2005, Rosenberg2006, Kole2007, Brigo2010, Meucci2011}, derivative pricing, see e.g., \cite{Rosenberg2003, Bennett2004, Cherubini2004, Goorbergh2005, Hull2006, Hofert2011}, and portfolio optimization, see e.g., \cite{Hennessy2002, Patton2004, DiClemente2004, Boubaker2013}.
For an overview of the literature on applications of copulas in finance the reader is referred to \cite{Genest2009, Patton2012b}. 
%The main advantage of the copula is that it captures the full statistical dependence between random variables eliminating the impact of the marginal distributions. 
%The dependence structure and the marginal distributions can be modeled separately and joined together resulting in new multivariate distributions with different behavior. 
Copulas allow to separate the dependence structure of random variables from their marginal distributions. This is sometimes useful in statistical applications as the dependence structure and the marginal distributions can be modeled separately and joined together resulting 
in new multivariate distributions with different behavior. For a discussion on difficulties in the application of copulas the reader is referred to \cite{Mikosch2006, Genest2006, Joe2006}. Here, we simply view the copulas as providing a standardized way for the study of statistical dependences. The marginal distributions are mapped to the uniform distribution;  the statistical dependence is considered in terms of the marginal cumulative distribution functions.

Recently, we identified market states as clusters of similar correlation matrices and studied their corresponding correlation structures \cite{Chetalova2015}. The correlation structure, however, does not capture the full 
statistical dependence between financial return time series. Here, we choose a copula approach to study the dependence structure of market states \cite{Marcel2015}.
To this end, we estimate empirical copulas for many stock return pairs and average over all of them to obtain an empirical pairwise copula for each market state. We stress that the identification of market states 
relies on the correlation matrices,
the copulas are used only for analyzing the states and not defining them. To estimate the empirical copulas we use both original and locally normalized returns.
%The original return time series exhibit non-stationary behavior, namely time-varying trends and volatilities \cite{black76, christie82, Schwert1989}. 
The original return time series exhibit time-varying trends and volatilities \cite{black76, christie82, Schwert1989}.
These have to be taken into account when transforming the marginal distributions to uniform ones.
To this end, we apply the method of local normalization \cite{Schaefer2010}, which leads to stationary time series while preserving the correlations between them. 
The resulting empirical copulas provide different information. The copulas for the original returns describe the dependence structure on a global scale, i.e., for the full time horizon, whereas the copulas for the locally normalized returns describe the dependence
structure on a local scale. %Furthermore, the copulas for the original returns exhibit stronger tail dependence, caused by periods with high volatility, and more pronounced asymmetry. }

The empirical pairwise copulas for each market state are compared with a bivariate K-copula, which arises from a random matrix approach introduced in \cite{Schmitt2013, Chetalova2013}. It models
the non-stationarity of true correlations by an ensemble of random matrices. The model yields a multivariate return distribution in terms of a modified Bessel function of the second kind, a so-called K-distribution. In \cite{Chetalova2015} the K-distribution
was found to provide a good description of the heavy-tailed empirical return distributions for each market state. 
Here, we aim to arrive at a consistent description within the random matrix model studying the agreement between K-copula and empirical dependence structure for each market state. 
In addition, our study provides further evidence for asymmetric dependencies between financial returns \cite{Longin2001, Ang2002, Hong2007}. We find an asymmetry in the tail dependence of empirical pairwise copulas which we study in 
more detail.

The paper is organized as follows. In section~\ref{section2} we review the basic concepts of copulas, stating the main result in the copula theory, the Sklar's theorem, which we use to derive a K-copula. 
In section~\ref{section3} we present the data set and recapitulate the identification of market states for the Nasdaq Composite market in the period $1992-2013$ as done in \cite{Chetalova2015}. In section~\ref{section4} we study the empirical copula densities for each market state and compare them with the K-copula. We conclude our findings in section~\ref{section5}.

\section{Copula}  \label{section2}

We begin with a short introduction to the concept of copulas in~\ref{section2.1}. For more details with an emphasis on the statistical and mathematical foundations of copulas the reader is referred to the textbooks of Joe \cite{Joe1997} and Nelsen \cite{Nelsen2006}. In~\ref{section2.2} we present the K-copula which plays a central role in this study. We restrict ourselves to the bivariate case, since later on we study empirical pairwise copulas.

\subsection{Basic Concepts} \label{section2.1}

Consider two random variables $X$ and $Y$. The joint distribution of $X$ and $Y$ contains all the statistical information about them. It can be expressed either in terms of the joint probability density function (pdf) 
$f_{X,Y}(x,y)$ or in terms of the joint cumulative distribution function (cdf) $ F_{X,Y}(x,y)$, where 
\begin{equation}
 F_{X,Y}(x,y)=\int\limits_{-\infty}^{x} {\rm d}x^{\prime} \int\limits_{-\infty}^{y} {\rm d}y^{\prime} \ f_{X,Y}(x^{\prime},y^{\prime})  \ .
\end{equation}
From the joint pdf $f_{X,Y}(x,y)$ one can extract the individual distributions of $X$ and $Y$ as follows
\begin{equation}
 f_X(x)=\int\limits_{-\infty}^{\infty} {\rm d}y \ f_{X,Y}(x,y) \ , 
\end{equation}
and analogously for $Y$. The densities $f_X(x)$ and $f_Y(y)$, called marginal probability density functions, and the corresponding marginal cumulative distribution functions  $F_X(x)$ and $F_Y(y)$ describe the individual statistical behavior of 
the random variables.  

When dealing with correlated random variables, one is interested in their statistical dependence. The Pearson correlation coefficient is commonly used as a measure of dependence. It is defined as
\begin{equation}
 C_{X,Y}=\frac{{\rm Cov}(X,Y)}{\sigma_X \sigma_Y} \ ,
 \label{pearson}
\end{equation}
where ${\rm Cov}(X,Y)$ is the covariance of both random variables and $\sigma_X$ and $\sigma_Y$ are the respective standard deviations. However, the correlation coefficient only measures the linear dependence between the random variables. 
%To capture the full statistical dependence one has to consider their copula. 

Copulas provide a natural way to study the statistical dependence of random variables. The transformation 
\begin{equation}
 U_i=F_i(i) \qquad i=X,Y
\end{equation}
leads to  new random variables with uniform distributions on the unit interval, called the rank of $X$ and $Y$, respectively. Their joint distribution is called a copula. 
It describes the dependence structure of the random variables $X$ and $Y$ separated from their marginal distributions.

A central result in the copula theory is Sklar's theorem which enables us to separate any multivariate distribution function into two components: the marginal distributions of each random variable and their copula
\begin{equation}
 F_{X,Y}(x,y)={\rm Cop}_{X,Y}(F_X(x),F_Y(y)) \ .
 \label{sklar}
\end{equation}
If the marginal distribution functions are continuous, the copula that satisfies equation~(\ref{sklar}) is given by
\begin{equation}
 {\rm Cop}_{X,Y}(u,v)=F_{X,Y}(F^{-1}_X(u),F^{-1}_Y(v)) \ , 
 \label{copula}
\end{equation}
where $F^{-1}_X$ and $F^{-1}_Y$ represent the inverse cumulative distribution functions, the so called quantile functions. This equation allows to extract the dependence structure directly from the joint distribution function.
From the copula (\ref{copula}) one can compute the copula density as follows
\begin{equation}
 {\rm cop}_{X,Y}(u,v)=\frac{\partial^2}{\partial u \partial v} {\rm Cop}_{X,Y}(u,v) \ .
 \label{cdensity}
\end{equation}

\subsection{K-copula} \label{section2.2}

The K-copula arises from a random matrix model introduced in \cite{Schmitt2013} to model time-varying correlations between financial time series \cite{Fenn2011, Muennix2012}. It was first used in \cite{Marcel2015}, where it was found to describe the empirical dependencies in financial data
much better than a Gaussian copula.

Consider a market consisting of $K$ stocks. At each time $t \left(t=1, \dots, T\right)$ we assume that the return vector $ r (t) = \left( r_1 (t),\dots ,r_K (t) \right)$ is drawn from a multivariate normal distribution with a covariance matrix $\Sigma_t$
\begin{equation}
g ( r| \Sigma_t) = \frac{ 1 }{ \sqrt{\det 2\pi  \Sigma_t } } \exp \left( - \frac{ 1 }{ 2 }  r^\dagger \Sigma^{-1}_t r \right)  \ ,
\label{eq:multidist}
\end{equation}
where we suppress the argument $t$ of $r$ to simplify our notation. We now model the time-dependent covariance matrix $\Sigma_t$ by a Wishart random matrix $AA^{\dagger}$, where the $K\times N$ model matrix $A$ is drawn from a 
Gaussian distribution with the pdf
%The time-dependent covariance matrix is modeled by a random Wishart matrix $AA^{\dagger}$. The $K\times N$ model matrix $A$ is drawn from a Gaussian distribution with the pdf
\begin{equation}
w ( A | \Sigma ,N) = \sqrt { \frac{ N }{ 2 \pi } }^{ K N } \frac{ 1 }{ \sqrt{\det \Sigma }^N } \exp \left( - \frac{ N }{ 2 } \tr  A^\dagger \Sigma^{-1}  A \right) \  .
\label{wishart}
\end{equation}
Here, $\Sigma$ represents the average covariance matrix estimated over the sample of all $r(t), t=1,\dots, T$.  %covariance matrix estimated %, computed over the considered sample. This sample may be any arbitrary time window or a market state. 
Hence, the time-dependent covariance matrices are modeled by an ensemble of Wishart matrices $AA^{\dagger}$ which fluctuate around the sample average $\Sigma$. Averaging the multivariate normal distribution~(8) with 
the random covariance matrix $AA^{\dagger}$
over the Wishart ensemble leads to a K-distribution for the multivariate returns 
\begin{equation}
\begin{split}
\langle  g \rangle ({r}| {\Sigma},N) &= \int  {\rm d}[ {A} ] \ w ({A}| {\Sigma},N) \ g (r| { A A^\dagger }) \\
&=\frac{1}{ (2 \pi)^K \Gamma( N/2 ) \sqrt{ \det \Sigma } } \int\limits_0^\infty {\rm d} z \ z^{\frac{ N }{ 2 } - 1 } {\rm e}^{-z} \sqrt{ \frac{ \pi N }{ z } }^K \exp \left( - \frac{ N }{ 4 z }  r^\dagger \Sigma^{-1} r \right)\\
& = \frac{ \sqrt{2}^{ 2-N } \sqrt{N}^K }{ \Gamma(N/2) \sqrt{ \det (2 \pi \Sigma) } } \frac{ \mathcal{K}_{\frac{K-N}{2}} \left( \sqrt{ N r^\dagger \Sigma^{-1} r} \right) }{ \sqrt{ N  r^\dagger \Sigma^{-1} r}^{\frac{K-N }{ 2 } } }  \ ,
\end{split}
\label{Kdistribution}
\end{equation}
where $\mathcal K_\nu$ is the modified Bessel function of the second kind of order $\nu = (K-N)/2$. It depends only on the average covariance matrix $\Sigma = \sigma C \sigma$ estimated over the whole sample, where $C$ is the average
correlation matrix and $\sigma={\rm diag}\left( \sigma_1, \dots, \sigma_K \right)$ the diagonal matrix of the standard deviations, and a free parameter $N$, 
%which characterizes the fluctuations around $\Sigma$.
which governs the variance of the Wishart ensemble
\begin{align}
 {\rm var} ([AA^\dagger]_{kl})=\frac{\Sigma_{kl}^2+\Sigma_{kk} \Sigma_{ll}}{N} \ ,
\end{align}
where $\Sigma_{kl}$ is the $kl$-th element of the average covariance matrix $\Sigma$. Thus, it characterizes the strength of fluctuations around the average covariance matrix in the considered sample. The larger $N$, the smaller the fluctuations around $\Sigma$, eventually vanishing in the limit
$N \rightarrow \infty$.

\begin{figure*}[t]
\centering
\includegraphics[width=\textwidth]{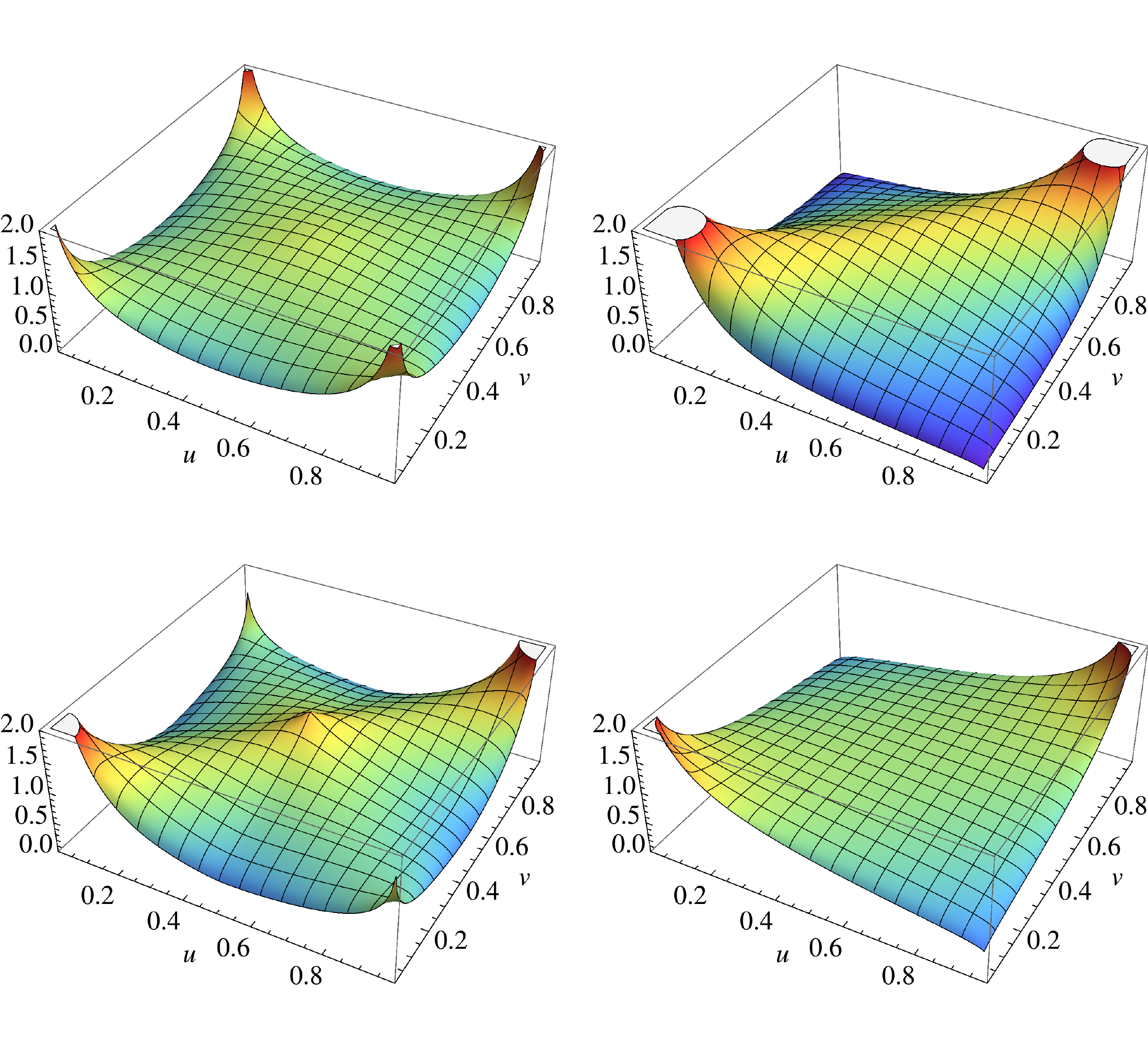}
\caption{ K-copula densities ${\rm cop}_{c,N}(u,v)$ for different parameter values. Top: $N=5$, (left) $c=0$ and (right) $c=0.5$ , bottom: $c=0.2$, (left) $N=3$ and (right) $N=30$.}
\label{fig1}
\end{figure*}

The K-copula is the dependence structure which arises for the K-distribution (\ref{Kdistribution}). For the bivariate case $K=2$, the pdf of the vector $r=(r_1,r_2)$ reads
\begin{equation}
f_{c,N}(r_1,r_2) =\langle  g \rangle ({r}| {\Sigma},N)= \frac{1}{\Gamma(N/2)} \int\limits_{0}^{\infty}  {\rm d}z \frac{z^{N/2-1}  {\rm e}^{-z}}{\sqrt{1-c^2}}\frac{N}{4 \pi z} \exp \left( -\frac{N}{4 z} \frac{r_1^2-2cr_1r_2+r_2^2}{1-c^2} \right) .
 \label{Kdist_twodim}
\end{equation}
Here, we used the covariance matrix
\begin{equation}
 \Sigma= \begin{pmatrix} \sigma_1^2 & \sigma_1 \sigma_2 c \\ \sigma_1 \sigma_2 c & \sigma_2^2  \end{pmatrix} = \begin{pmatrix} 1 & c \\ c & 1  \end{pmatrix} \ ,
\end{equation}
where $c$ denotes the average correlation coefficient, estimated over the whole sample. We chose the standard deviations one, $\sigma_1=\sigma_2=1$, since the copula is independent of the marginal distributions. 
Then, the marginal distribution densities are identical, $f_1(r_1)=f_2(r_2)$,  where
\begin{equation}
 f_1(r_1)=\int\limits_{-\infty}^{\infty} {\rm d}r_2 \ f_{c,N}(r_1,r_2) =\frac{1}{\Gamma(N/2)} \int\limits_{0}^{\infty}  {\rm d}z \ z^{N/2-1}  {\rm e}^{-z} \sqrt{\frac{N}{4 \pi z}}  \exp \left( -\frac{N}{4 z} r_1^2 \right) \ .
 \label{marginals}
 \end{equation}
According to equation~(\ref{copula}) the bivariate K-copula is given by
\begin{equation}
  {\rm Cop}_{c,N}(u,v)=F_{c,N}(F^{-1}_1(u),F^{-1}_2(v))  \ ,
  \label{Kcopula}
\end{equation}
where $c$ and $N$ are the parameters of the copula, $F_{c,N}$ denotes the cumulative distribution function of the bivariate distribution (\ref{Kdist_twodim}), and $F^{-1}$ the inverse distribution function of the marginal cdf  given by
\begin{align}
%\begin{split}
F_{c,N} (r_1,r_2)&= \int\limits_{-\infty}^{r_1}  {\rm d}\xi \int\limits_{-\infty}^{r_2}  {\rm d}\zeta \ f_{c,N}(\xi,\zeta) \notag \\
&=\int\limits_{-\infty}^{r_1}  {\rm d}\xi \int\limits_{-\infty}^{r_2}  {\rm d}\zeta \int\limits_{0}^{\infty} \frac{{\rm d}z}{\Gamma(N/2)} \frac{z^{N/2-1} {\rm e}^{-z}}{\sqrt{1-c^2}}\frac{N}{4 \pi z}  \exp \left( -\frac{N}{4 z} \frac{\xi^2-2c\xi \zeta+\zeta^2}{1-c^2} \right) \notag \\
&=\int\limits_{-\infty}^{r_1}  {\rm d}\xi \int\limits_{-\infty}^{r_2} {\rm d}\zeta \ \frac{N \sqrt{\frac{N (\xi^2-2c\xi \zeta+\zeta^2)}{1-c^2}}^{\frac{N-2}{2}} }{\pi \ \Gamma(N/2) \sqrt{2}^{N}\sqrt{1-c^2}} \  \mathcal K_{\frac{2-N}{2}}\left( \sqrt{\frac{N (\xi^2-2c\xi \zeta+\zeta^2)}{1-c^2}}\right) \notag \\
%\end{split}
\end{align}
and 
\begin{align}
%\begin{split}
F_1(r_1)&=\int\limits_{-\infty}^{r_1}  {\rm d}\xi \  f_1(\xi)=  \int\limits_{-\infty}^{r_1}  {\rm d}\xi \frac{1}{\Gamma(N/2)} \int\limits_{0}^{\infty}  {\rm d}z \ z^{N/2-1}  {\rm e}^{-z}\sqrt{\frac{N}{4 \pi z}}  \exp \left( -\frac{N}{4 z} \xi^2 \right) \notag \\
&=\int\limits_{-\infty}^{r_1}  {\rm d}\xi \ \frac{\sqrt{N}\sqrt{N\xi^2}^{\frac{N-1}{2}}}{\sqrt{\pi} \ \Gamma(N/2) \sqrt{2}^{N-1}} \ \mathcal K_{\frac{1-N}{2}} \left( \sqrt{N \xi^2}\right) \ .
%\end{split}
\end{align}
The K-copula density can be obtained from the K-copula (\ref{Kcopula}) by differentiation
\begin{equation}
  {\rm cop}_{c,N}(u,v)= \frac{\partial^2  {\rm Cop}_{c,N}(u,v) }{\partial u \partial v} = \frac{f_{c,N}(F_1^{-1}(u),F_2^{-1}(v))}{f_1(F_1^{-1}(u)) f_2(F_2^{-1}(v))} \ .
\end{equation}
It depends only on the average correlation $c$ and the free parameter $N$, which characterizes the fluctuations around $\Sigma$. Figure~\ref{fig1} shows the K-copula density for different parameter values. 
The stronger the average correlation $c$ and the lower the parameter $N$, the higher is the probability for extreme co-movements.

Furthermore, we note that the K-copula is a symmetric copula. It is based on the elliptical distribution (\ref{Kdistribution}) and thus it belongs to the class of elliptical copulas.

\section{Identification of market states}   \label{section3}

We now present the data set and identify market states as clusters of correlation matrices with similar correlation structures, as done in \cite{Chetalova2015}. These market states will be the object of our empirical study in the next section. %, where we identified market states.% will be the object of our study in the next section.

We consider $K=258$ stocks of the Nasdaq Composite index traded in the 22-year period from January 1992 to December 2013 \cite{Yahoo}  which corresponds to  5542 trading days. For each stock $k$ we calculate the return time series
\begin{equation}
 r_k (t) = \frac{ S_k(t + \Delta t)-S_k(t) }{ S_k(t) } \ , \quad k=1,\dots, K \ ,
\label{returns}
\end{equation}
where $ S_k(t) $ is the price of the $ k $-th stock at time $ t $ and $ \Delta t $ is the return interval which we chose to be one trading day. 

%Return time series are non-stationary. In particular, drift and volatilities  fluctuate considerably in time. Since the correlation coefficient (\ref{pearson}) averages over time-varying trends and volatilities, the non-stationarity results in an estimation error of the correlations. 
Empirical return time series exhibit time-varying drift and volatilities. The correlation coefficient~(\ref{pearson}) averages over these time-dependent parameters which results in an estimation error of the correlations.
In order to eliminate this kind of error we employ the method of local normalization \cite{Schaefer2010}.
For each return time series $k$ we subtract the local mean and divide by the local standard deviation
\begin{equation}
\hat r_k (t) = \frac{ r(t)-\langle r(t)\rangle_n }{ \sqrt{\langle r^2(t) \rangle_n - \langle r(t)\rangle^2_n} } \quad ,
\label{normreturns}
\end{equation}
where $\langle \dots \rangle_n$ denotes the local average which runs over the $n$ most recent sampling points. For daily data we use $n=13$ as discussed in \cite{Schaefer2010}. 
The local normalization removes the local trends and variable volatilities while preserving the correlations between the time series.

\begin{figure*}[h]
\centering
\includegraphics[width=\textwidth]{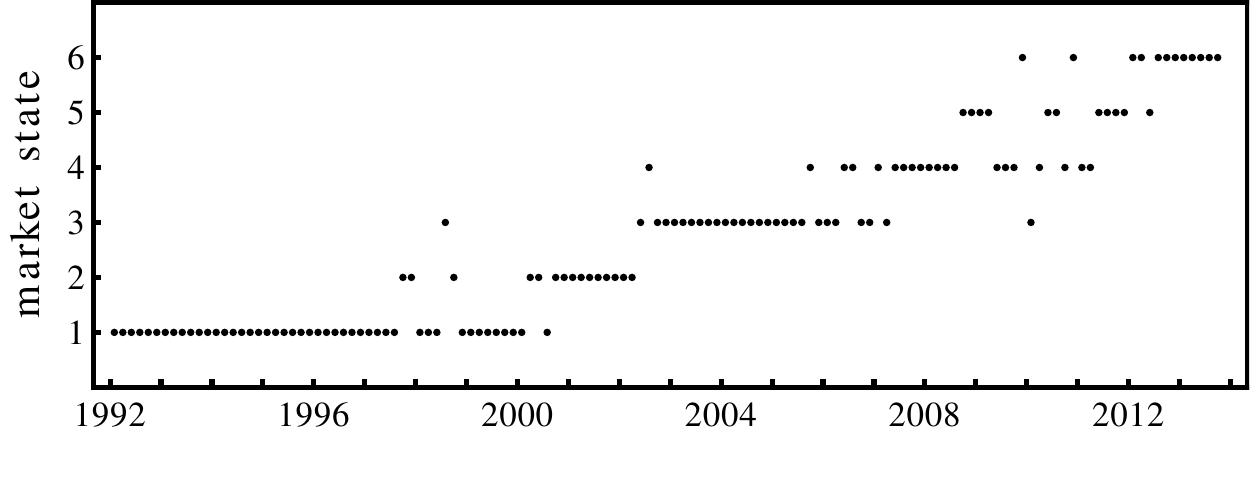}
\caption{ Time evolution of the market in the observation period $1992-2013$. Each point represents a correlation matrix measured over a two-month time window. }
\label{fig2}
\end{figure*}

Using the locally normalized daily returns we now obtain a set of 131 correlation matrices estimated on disjoint two-month intervals of the 22-year observation period. 
To identify market states we perform a clustering analysis based on the partitioning around medoids algorithm \cite{Kaufman1990}, where the number of clusters is estimated via the gap statistic \cite{Tibs2001}.
The clustering analysis separates the set of 131 correlation matrices into six groups based on the similarity of their correlation structures. Each group is associated with a market state. 
We point out that the identification of market states is performed ex-post, the clustering algorithm  has the correlation matrices of all times.
Figure~\ref{fig2} shows the time evolution of the six  market states. We observe that the market switches back and forth between states. Sometimes it remains in a state for a long time, sometimes it jumps briefly to another state and 
returns back or evolves further. On longer time scales the market evolves towards new states, whereas previous states die out.

In the next section we will study the statistical dependence for each market state. To obtain the return time series for each state, we proceed as follows: We take the complete return time series, $r(t)$ or $\hat r(t)$, 
and divide it into a sequence of disjoint two month intervals. We merge all intervals belonging to a given state according to the cluster analysis.
We note that the return time series for the six market states differ in length. 

\section{Empirical results}  \label{section4}

We present the empirical pairwise copula densities for each market states in~\ref{section4.1} and compare them with the K-copula densities in~\ref{section4.2}. In~\ref{section4.3} we study the asymmetry of the tail dependence 
of the empirical copula densities in more detail.

\subsection{Empirical pairwise copulas for each market state} \label{section4.1}

To estimate the empirical pairwise copula of two return time series $r_k(t)$ and $r_l(t)$, we first have to transform them into uniformly distributed time series. To achieve this, we employ the empirical distribution function
\begin{equation}
u_k(t)=F_k(r_k(t))= \frac{1}{T} \sum_{\tau=1}^T {\bf 1} \{ r_k(\tau) \leq r_k(t)\} -\frac{1}{2T} \ ,
\end{equation}
where ${\bf 1}$ is the indicator function, $T$ denotes the length of the time series and the factor $1/2$ ensures that the values of the transformed time series $u_k(t)$ lie in the interval $(0,1)$.
The empirical copula density of the time series $r_k(t)$ and $r_l(t)$ is then the two-dimensional histogram of the transformed time series $u=u_k(t)$ and $v=u_l(t)$.

An accurate estimation of the copula density requires a large amount of data. Thus, for each state we compute the copula densities of all $K(K-1)/2$ stock pairs as two-dimensional histograms of the transformed time series and then average over all pairs
\begin{equation}
{\rm cop}^{(i)}(u,v)=\frac{2}{K(K-1)} \sum_{k=1}^{K-1} \sum_{l=k+1}^{K} {\rm cop}^{(i)}_{k,l}(u,v) \ , \qquad i=1,\dots,6\ ,
\end{equation}
where the upper index $i$ denotes the state number. For the bin size of the histograms we choose $\Delta u=\Delta v =0.05$. We note that as the length of the time series for each market state is rather short, we are not able to study 
the full copulas for each stock pair $k,l$ separately. Hence, we cannot make any direct statements about similarity and dispersion regarding the full dependence structure. Our results only yield statements about the empirical dependence 
structure on average. 

\begin{figure*}
\centering
\includegraphics[width=\textwidth]{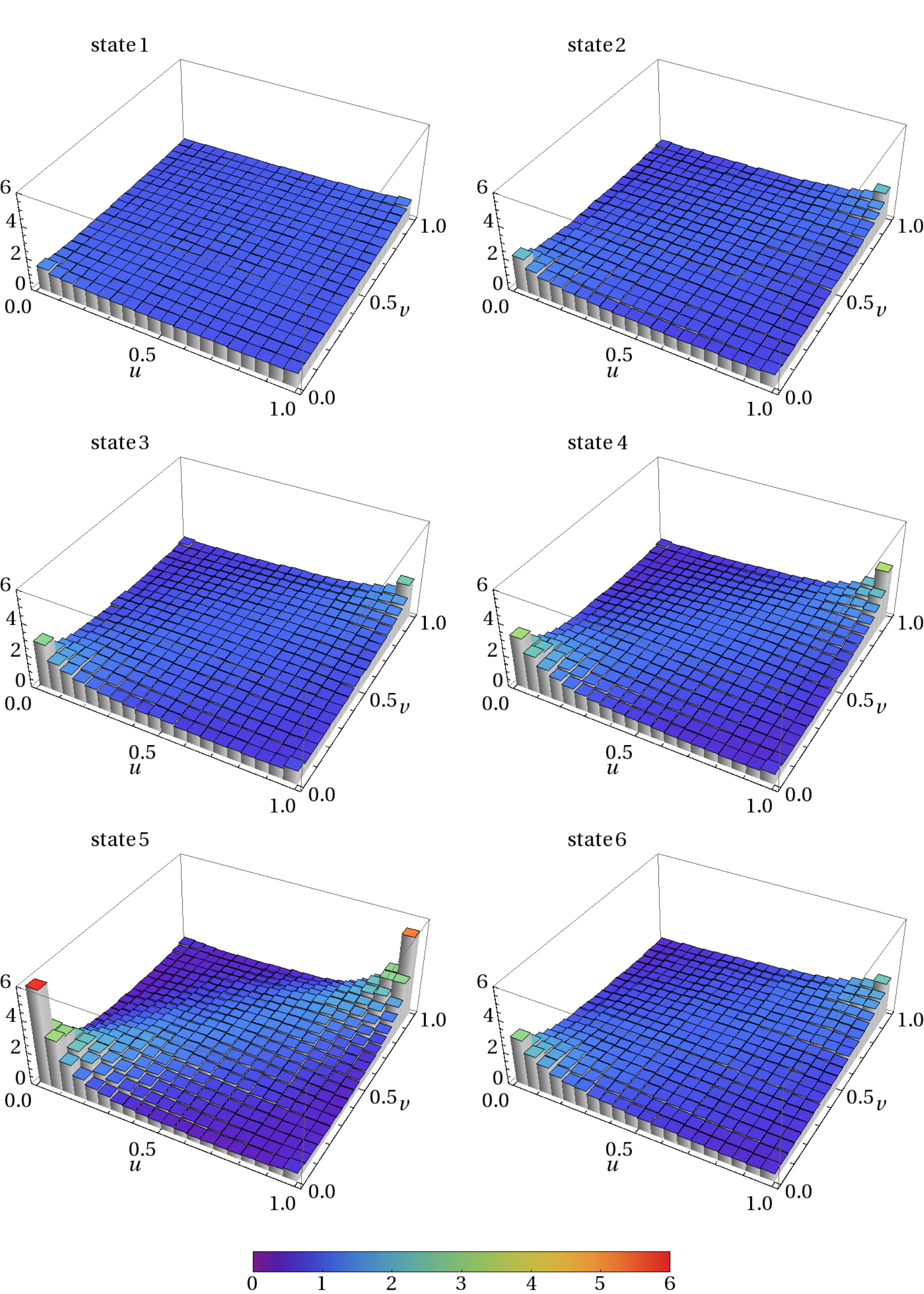}
\caption{ Empirical pairwise copula density ${\rm cop}^{(i)}(u,v)$ for the original returns. }
\label{fig3}
\end{figure*}

Figure~\ref{fig3} shows the empirical pairwise copula densities for the original returns (\ref{returns}) for each of the six market states. 
We observe a variation of the dependence structure from state to state, particularly visible in the tails. In state 1, which covers the period from 1992 to roughly 2000, we find a rather flat copula density, indicating low 
dependence between return pairs.
In state 2 we observe deviations from the flat copula density particularly in the tails, which become more and more pronounced in state 3 and 4. State 5, first appearing during the financial crisis in 2008, exhibits the strongest dependence.
The dependence decreases again in state 6. 

\begin{figure*}
\centering
\includegraphics[width=\textwidth]{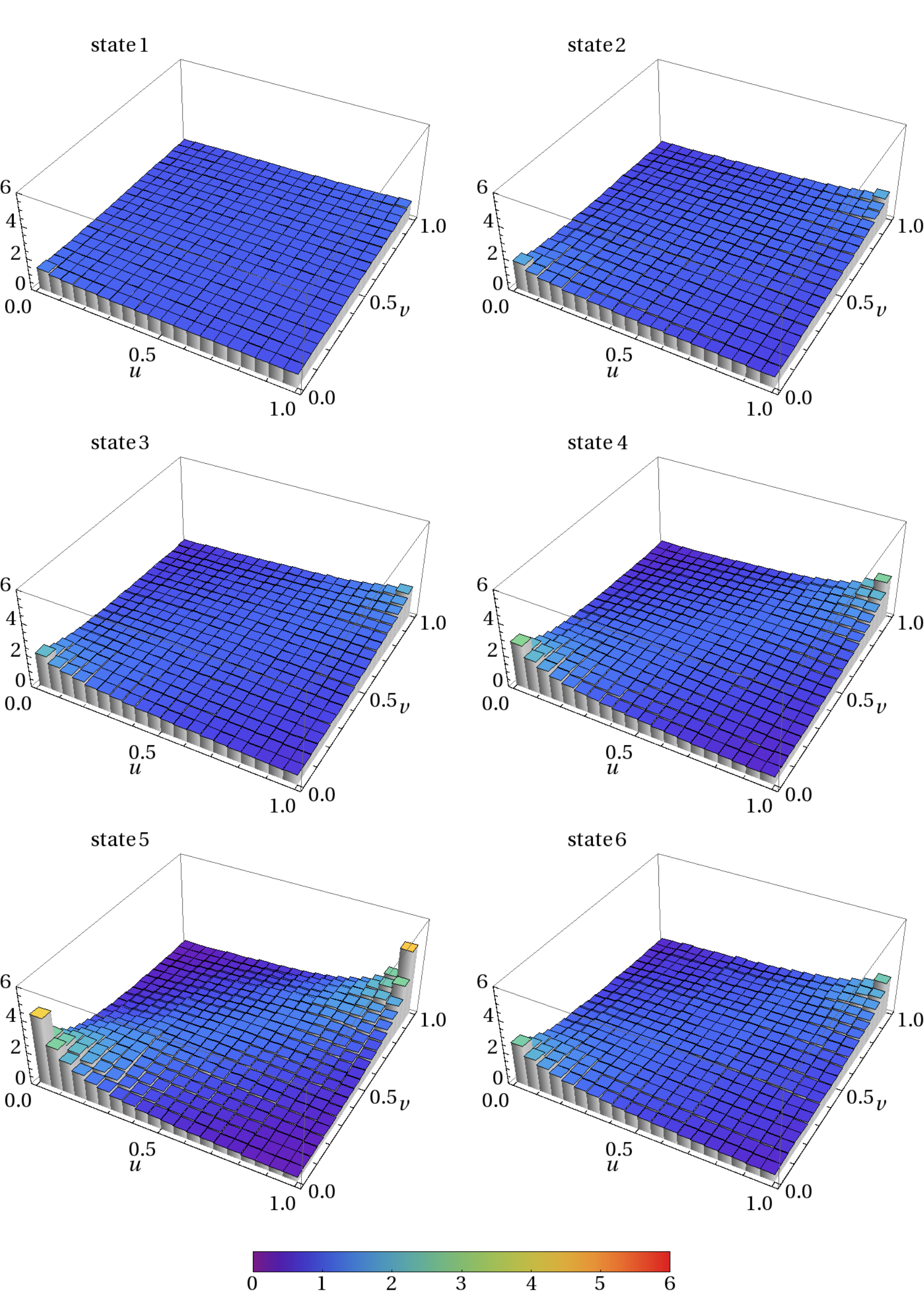}
\caption{ Empirical pairwise copula density ${\rm cop}^{(i)}(u,v)$ for the locally normalized returns. }
\label{fig4}
\end{figure*}

Figure~\ref{fig4} shows the empirical pairwise copula densities for the locally normalized returns (\ref{normreturns}). The dependence structures of the six states are mostly preserved after performing local normalization.
Deviations are observed in the lower-left and the upper-right corners where the copula densities for the original returns exhibit higher peaks. The reason for that is the time-varying volatility. 
During periods of high volatility stocks tend to have large returns which contribute to the corners of the copula density. On the other hand, in periods of high volatility the correlations between stocks become stronger. This leads to
higher peaks in the corners of the dependence structure for the original returns. 

It is important to note that the copulas for the original and the locally normalized returns contain different statements. The copulas for the original returns describe the dependence structure for the full time horizon. On the other hand, the copulas for the locally normalized returns provide information about the statistical dependence on a local scale. 

Furthermore, we observe that the empirical copula densities are asymmetric with respect to opposite corners. We find stronger dependence in the lower tail than in the upper one, that is, the dependence between large negative returns is stronger than  the dependence between large positive ones.
This asymmetry is an important feature of empirical copula densities and thus we will discuss it in more details in section~\ref{section4.3}.


\subsection{Comparison with the K-copula} \label{section4.2} 

\begin{figure*}
\centering
\includegraphics[width=\textwidth]{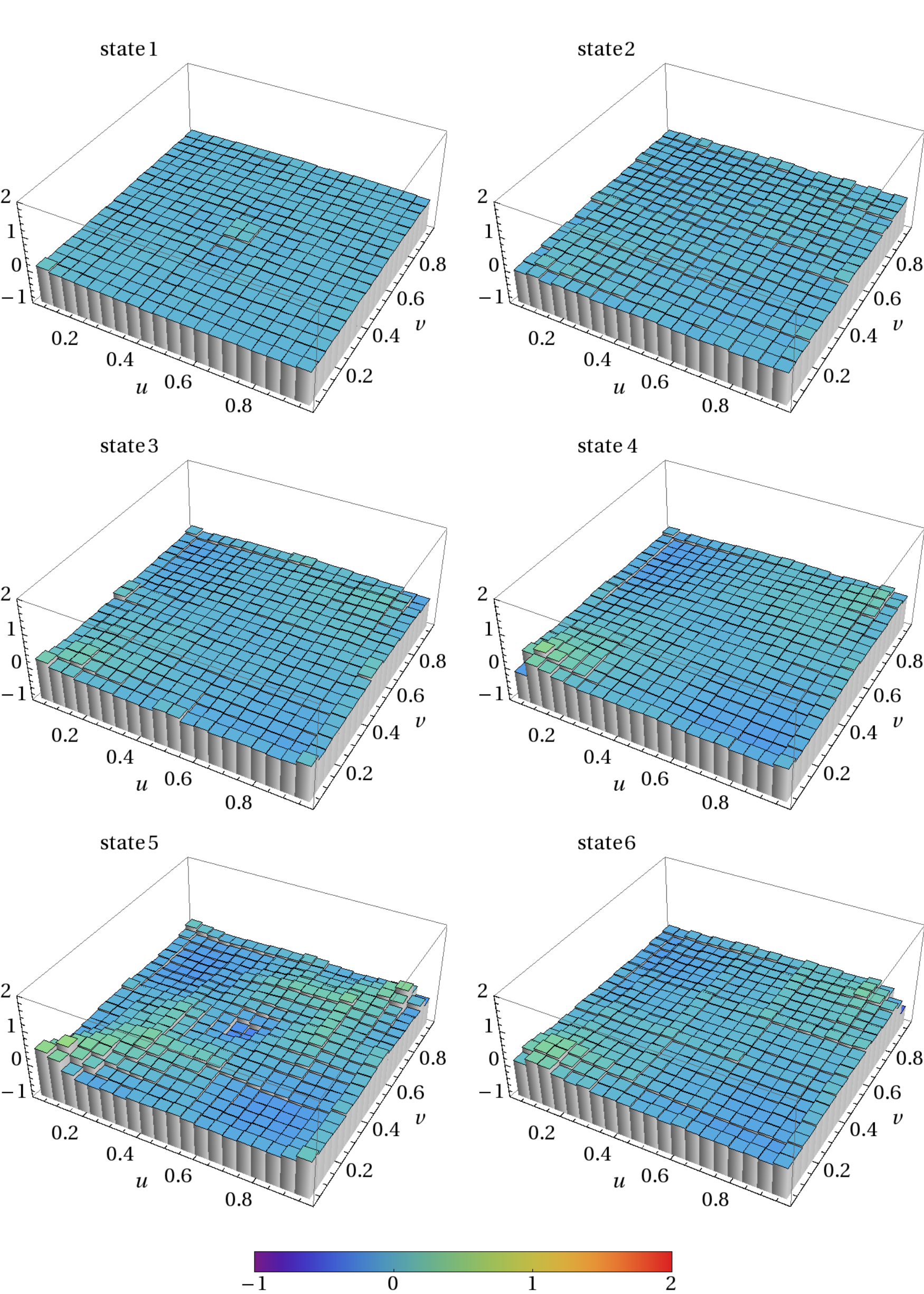}
\caption{ Difference between the empirical copula density and the K-copula density ${\rm cop}^{(i)}(u,v)-{\rm cop}^{(i)}_{\bar{c},N}(u,v)$ for the original returns. }
\label{fig5}
\end{figure*}

\begin{figure*}
\centering
\includegraphics[width=\textwidth]{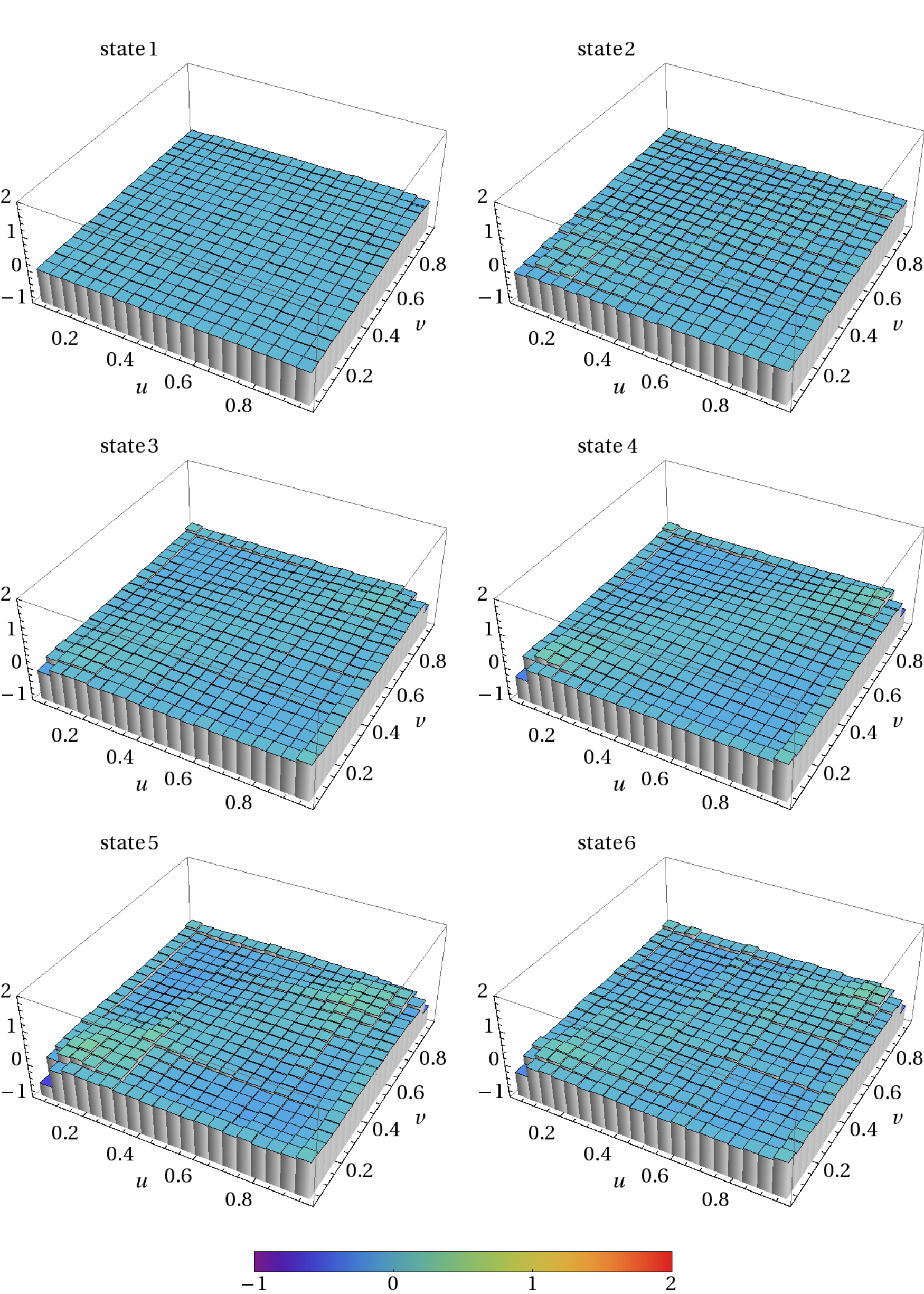}
\caption{ Difference between the empirical copula density and the K-copula density ${\rm cop}^{(i)}(u,v)-{\rm cop}^{(i)}_{\bar{c},N}(u,v)$ for the locally normalized returns. }
\label{fig6}
\end{figure*}

We now compare the empirical pairwise copula densities for each market state with the K-copula densities. Again, we consider both the original and the locally normalized returns.
The K-copula density is obtained in the following way: We calculate the K-copula according to equation (\ref{Kcopula}) where the integrals are performed numerically. The K-copula density for each bin of size $\Delta u=\Delta v =0.05$ is 
then estimated by
\begin{equation}
\begin{split}
 {\rm cop}_{\bar{c},N}(u,v)= {\rm Cop}_{\bar{c},N}(u,v)-{\rm Cop}_{\bar{c},N}(u,v-\Delta v)&-{\rm Cop}_{\bar{c},N}(u-\Delta u,v)\\ &+{\rm Cop}_{\bar{c},N}(u-\Delta u,v-\Delta v) \ .
 \end{split}
\end{equation}
For comparison we compute the difference between empirical and analytical 
copula density for each state
\begin{equation}
 {\rm cop}^{(i)}(u,v)-{\rm cop}^{(i)}_{\bar{c},N}(u,v) \ ,  \qquad i=1,\dots,6 \ ,
\end{equation}
where $\bar c$ is the average correlation coefficient of all $K(K-1)/2$ stock pairs for the considered state. The free parameter $N$ is estimated by a fit which minimizes the mean squared difference between empirical and analytical copula density. 
The parameter values for each market state are summarized in table~\ref{table1}.
\begin{table}
\centering
\begin{tabular}{rrrrrrrr}
\hline
    returns & & state 1 & state 2 & state 3 & state 4 & state 5 & state 6\\
\hline
\multirow{2}{*}{original} 
& $\bar{c}$     &   0.046  &  0.13   & 0.17 & 0.25 & 0.42 & 0.22 \\[2mm]
& $N$          &  41.7 &  11.7 & 8.4 & 5.6 & 2.8     & 10.0       \\
\hline\\[0.5mm]
\hline 
\multirow{2}{*}{loc. normalized} 
&$\bar{c}$     &   0.048  & 0.13    &  0.19 & 0.28 & 0.43 & 0.25 \\[2mm]
& $N$         &   70.7  & 28.6    &  29.8 & 15.8 & 7.4 & 20.4       \\
\hline
\end{tabular}
\caption{Parameters of the K-copula density for the original and the locally normalized returns.}
\label{table1}
\end{table}
The differences between the empirical copula density and the K-copula density for each state are presented in figure~\ref{fig5} for the original and in figure~\ref{fig6} for the locally normalized returns.
Overall, we find a good agreement between empirical and analytical copula densities. The K-copula seems to capture the dependence structure of the first three states very well. Small deviations from the K-copula density are observed for state 4 and 6. Only the dependence structure of state 5 cannot be captured by the K-copula.
For the locally normalized returns we find a better agreement, 
which is reflected in the smaller mean squared differences, see table~\ref{table2}. Deviations are observed mainly in the corners of the copula densities. The empirical copula densities exhibit stronger dependence in the tails, i.e., the probability 
for extreme co-movements is underestimated by the K-copula.
\begin{table}
\centering
\begin{tabular}{rrrrrrr}
\hline
  returns  & state 1 & state 2 & state 3 & state 4 & state 5 & state 6\\
\hline
original       &  0.11   &  0.41   & 1.22 & 2.47 & 5.53 & 2.84 \\[2mm]
loc. normalized  & 0.047       & 0.38  & 0.70 & 1.43 & 2.94  & 1.50       \\
\hline
\end{tabular}
\caption{Mean squared differences between empirical and K-copula densities.}
\label{table2}
\end{table}

It is important to note that the K-copula captures the empirical dependence structure rather well with only one free parameter. 
%It outperforms other elliptical copulas like the Gaussian or the $t$ copula. 
However, due to its symmetric nature it cannot account for the asymmetry observed in the data. 
The skewed Student's $t$-copula is an alternative proposed by \cite{Demarta2005} which was found to account for asymmetric dependencies in financial data \cite{Sun2008,Ammann2009}. 
It captures the empirical dependence structure of the original returns better than the K-copula due to the presence of an additional parameter which accounts for the asymmetry \cite{Marcel2015}. Nevertheless, 
here we confine ourselves to the comparison of the empirical copulas with the K-copula as we aim to arrive at a consistent picture within the random matrix model. We discuss 
the asymmetry in more detail in the following.

\subsection{Asymmetry of the tail dependence}   \label{section4.3} 

\begin{figure*}
\centering
\includegraphics[width=\textwidth]{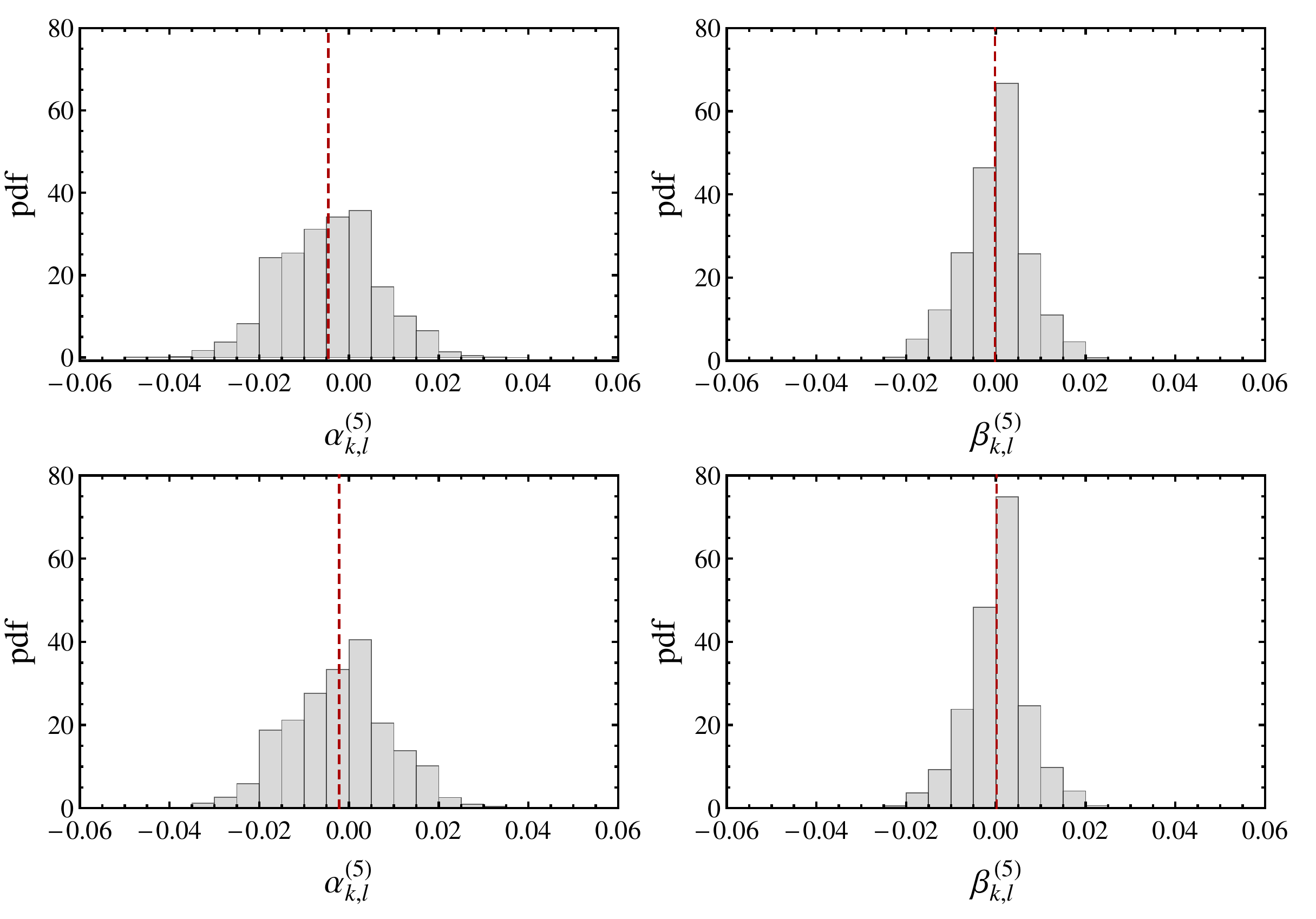}
\caption{ Histograms of the asymmetry values for all stock pairs $k$,$l$, exemplarily for state 5. Left: asymmetry for the positive tail dependence $\alpha^{(5)}_{k,l}$, right: asymmetry for the negative tail dependence: $\beta^{(5)}_{k,l}$. Top: for returns, bottom: for locally normalized returns.
The dashed red lines represent the corresponding mean values.}
\label{fig8}
\end{figure*}

Asymmetric dependence between returns has been reported by several authors, see e.g., \cite{Longin2001, Ang2002, Hong2007}. Our study provides further evidence revealing  a stronger lower tail dependence in the empirical copula densities
of market states. We now take a closer look at this asymmetry.
To this end, we estimate the tail dependence in the four corners for all $K(K-1)/2$ empirical pairwise copulas according to 
\begin{equation}
\begin{split}
 {\rm LL}^{(i)}_{k,l}&=\int \limits_0^{0.2} {\rm d}u \int \limits_0^{0.2} {\rm d}v \ {\rm cop}^{(i)}_{k,l}(u,v) \ , \\
 {\rm UL}^{(i)}_{k,l}&=\int \limits_{0.8}^1 {\rm d}u \int \limits_0^{0.2} {\rm d}v \ {\rm cop}^{(i)}_{k,l}(u,v) \ , \\
 {\rm UU}^{(i)}_{k,l}&=\int \limits_{0.8}^1 {\rm d}u \int \limits_{0.8}^{1} {\rm d}v \ {\rm cop}^{(i)}_{k,l}(u,v) \ , \\
 {\rm LU}^{(i)}_{k,l}&=\int \limits_{0}^{0.2} {\rm d}u \int \limits_{0.8}^{1} {\rm d}v \ {\rm cop}^{(i)}_{k,l}(u,v) \ , %\qquad i=1,\dots,6 \ .
\end{split}
\end{equation}
where the upper index $i=1, \dots, 6$ denotes the state number and the lower indices represent a stock pair $k,l$. Here, LL and UU refer to the lower-lower and upper-upper corners, respectively, and represent the positive
tail dependence, whereas UL and LU refer to the upper-lower and lower-upper corners, respectively, and represent the negative tail dependence.
%Here, LL refers to the lower-lower, UL to the upper-lower, UU to the upper-upper, and LU to the lower-upper corner. 
The asymmetry in the tail dependence can now be quantified by the differences
\begin{equation}
\begin{split}
 \alpha^{(i)}_{k,l}&= {\rm UU}^{(i)}_{k,l} - {\rm LL}^{(i)}_{k,l} \ , \\
 \beta^{(i)}_{k,l}&= {\rm LU}^{(i)}_{k,l} - {\rm UL}^{(i)}_{k,l} \ , \qquad i=1,\dots,6 \ ,
\end{split}
\end{equation}
where $\alpha^{(i)}_{k,l}$ captures the asymmetry of the positive and $\beta^{(i)}_{k,l}$ the asymmetry of the negative tail dependence for each stock pair $k$,$l$. Figure~\ref{fig8} shows the histograms of the asymmetry values exemplarily for state 5. 
For the returns we find a negative offset for the values of $\alpha^{(5)}_{k,l}$, whereas the values of $\beta^{(5)}_{k,l}$ are centered around zero. 
This indicates, on average, an asymmetry in the positive tail dependence, i.e., simultaneous large negative returns are more likely to occur than  simultaneous large positive returns. On the other hand, we do not find such an asymmetry 
in the negative tail dependence. For the locally normalized returns, we find much weaker asymmetry in the positive tail dependence and once again no asymmetry for the negative tail dependence. We note that the means of the asymmetry values are the relevant quantities, the 
asymmetry values for each pair are distributed around the mean due to statistical fluctuations. The standard deviation for $\alpha^{(5)}_{k,l}$ is 0.01 and for $\beta^{(5)}_{k,l}$ 0.007. Indeed, the asymmetry effect is very small. 
Still, it is clearly visible, see figure~\ref{fig3}.

\begin{figure*}
\centering
\includegraphics[width=\textwidth]{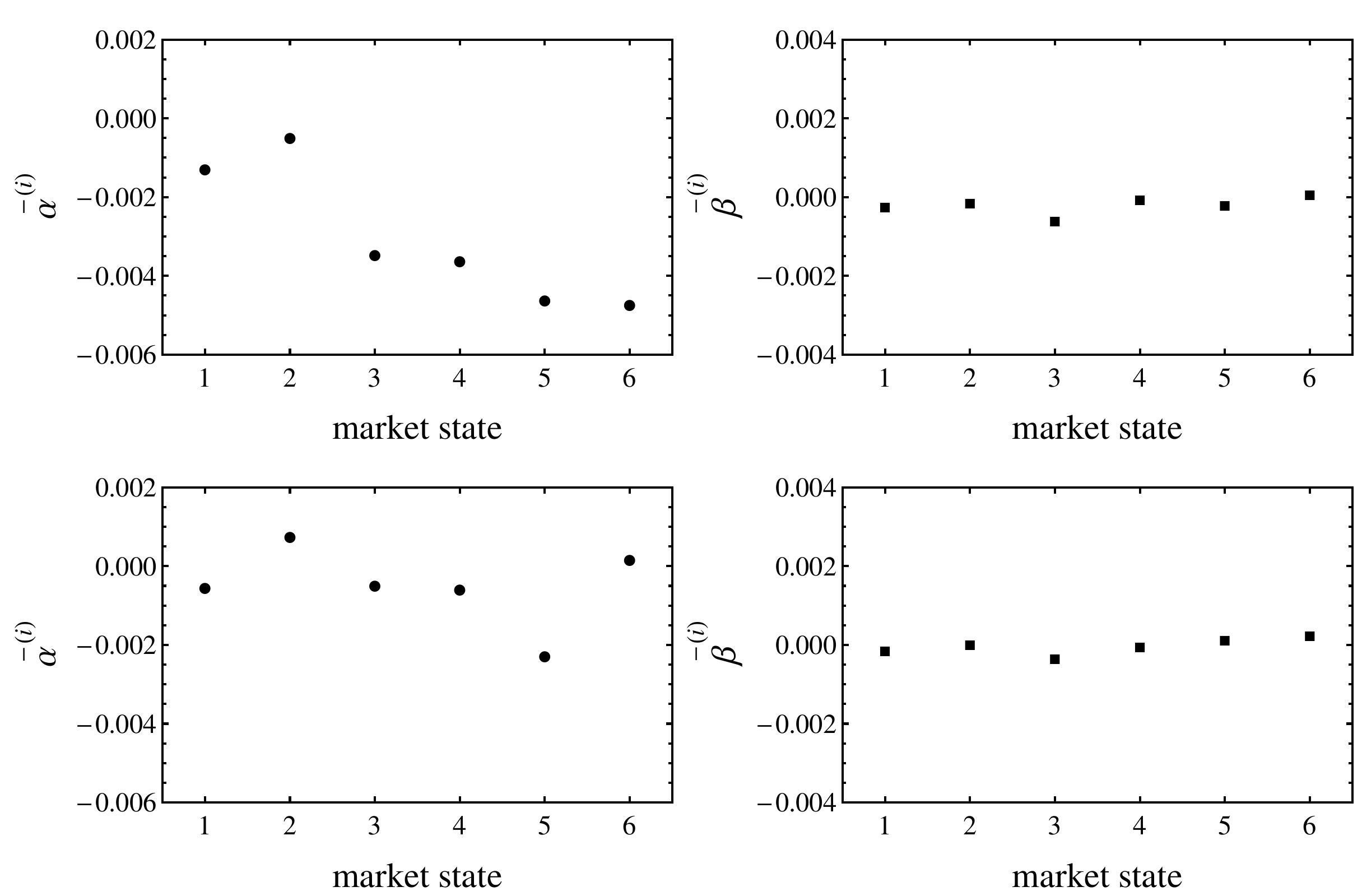}
\caption{ The mean asymmetry values for each state. Left: asymmetry for the positive tail dependence $\bar{\alpha}^{(i)}$, right: asymmetry for the negative tail dependence: $\bar{\beta}^{(i)}$. Top: for returns, bottom: for locally normalized returns.}
\label{fig9}
\end{figure*}

In the following, we study the asymmetry values for each market state.
Figure~\ref{fig9} shows the mean asymmetry values $\bar \alpha^{(i)}$ for each market state, obtained by averaging over all $\alpha^{(i)}_{k,l}$ for a given state. On a local scale the asymmetry in the positive tail dependence is much weaker.
Only state 5 still exhibits a certain amount of asymmetry. On the other hand, the asymmetry in the negative tail dependence is negligibly small for both original and locally normalized returns.

\section{Conclusion}  \label{section5}

We studied the dependence structure of market states by means of a copula approach. To this end, we estimated the empirical pairwise copulas for each state and compared them with the bivariate K-copula. 
The bivariate K-copula arises from a random matrix model where the non-stationarity of correlations is taken into account by an ensemble of random matrices. It is a symmetric, elliptical copula, which depends on two parameters:
the average correlation coefficient, estimated over the considered sample of returns, and a free parameter which characterizes the fluctuations around the average correlation in this sample.
We estimated the empirical pairwise copulas for both original and locally normalized returns. The local normalization removes the time-varying trends and volatilities while preserving the dependence structure. The corresponding copula 
describes the dependence structure on a local scale, whereas the copula of the original returns provides information about the dependence structure on a global scale, i.e., for the full time horizon. 
Overall, the K-copula captures the empirical dependence structure of market states. We found a good agreement, in particular for the copulas estimated on a local scale. %The agreement is even better for the locally normalized returns.
Thus, we obtain a consistent description within the random matrix model: The K-distribution describes the heavy-tailed return distributions while the K-copula captures the corresponding dependence structure. 
However, we found an asymmetry in the positive tail dependence, i.e., a stronger lower tail dependence, indicating a larger probability for simultaneous 
extreme negative returns. This asymmetry cannot be captured by our model.  It is more pronounced on a global scale.
On a local scale we find a much weaker asymmetry. However, in times of crisis the asymmetry is still clearly present. 
%In addition, our study provides further evidence for asymmetric dependencies between financial returns. 
%We find a stronger lower tail dependence, indicating a larger probability for simultaneous 
%extreme negative returns. On a local scale we find a much weaker asymmetry. However, in times of crisis the asymmetry is still clearly present. 

\section*{References}

%\bibliographystyle{unsrt}
%\bibliography{references}

\end{document}